\documentclass[aps,prx,twocolumn,superscriptaddress,floatfix]{revtex4-1}

\usepackage[latin9]{inputenc}
\usepackage{graphicx}
\usepackage{dcolumn}
\usepackage{bm}
\usepackage{amssymb}
\usepackage{microtype}
\usepackage{xfrac}
\usepackage{makecell}
\usepackage{array}
\usepackage[version=4]{mhchem}
\usepackage{gensymb}
\usepackage{multirow}
\usepackage[colorlinks=true]{hyperref}

\begin{document}

\title{First Order Preemptive Ising-nematic Transition in K$_{5}$Fe$_{4}$Ag$_{6}$Te$_{10}$}% Force line breaks with \\

\author{N.~Giles-Donovan}
\affiliation{Department of Physics, University of California, Berkeley, CA 94720, USA}
\affiliation{Material Sciences Division, Lawrence Berkeley National Lab, Berkeley, California 94720, USA}

\author{Y.~Chen}
\affiliation{Department of Physics, University of California, Berkeley, CA 94720, USA}
\affiliation{Material Sciences Division, Lawrence Berkeley National Lab, Berkeley, California 94720, USA}

\author{H.~Fukui}
\affiliation{Precision Spectroscopy Division, CSRR, SPring-8/JASRI, 1-1-1 Kouto, Sayo, Hyogo, 679-5198, Japan}
%\affiliation{Materials Dynamics Laboratory, RIKEN SPring-8 Center, Sayo, Hyogo 679-5148, Japan}

\author{T.~Manjo}
\affiliation{Precision Spectroscopy Division, CSRR, SPring-8/JASRI, 1-1-1 Kouto, Sayo, Hyogo, 679-5198, Japan}
%\affiliation{Materials Dynamics Laboratory, RIKEN SPring-8 Center, Sayo, Hyogo 679-5148, Japan}

\author{D.~Ishikawa}
\affiliation{Precision Spectroscopy Division, CSRR, SPring-8/JASRI, 1-1-1 Kouto, Sayo, Hyogo, 679-5198, Japan}
\affiliation{Materials Dynamics Laboratory, RIKEN SPring-8 Center, Sayo, Hyogo 679-5148, Japan}

\author{A.~Q.~R.~Baron}
\affiliation{Materials Dynamics Laboratory, RIKEN SPring-8 Center, Sayo, Hyogo 679-5148, Japan}
\affiliation{Precision Spectroscopy Division, CSRR, SPring-8/JASRI, 1-1-1 Kouto, Sayo, Hyogo, 679-5198, Japan}

\author{S.~Chi}
\affiliation{Neutron Scattering Division, Oak Ridge National Laboratory, Oak Ridge, TN, 37831, USA}

\author{H.~Zhong}
\affiliation{Center for Advanced Quantum Studies, School of Physics and Astronomy, Beijing Normal University, Beijing 100875, China}

\author{S.~Cao}
\author{Y.~Tang}
\author{Y.~Wang}
\affiliation{Center for Correlated Matter and School of Physics, Zhejiang University, Hangzhou 310058, China}

\author{X.~Lu}
\affiliation{Center for Advanced Quantum Studies, School of Physics and Astronomy, Beijing Normal University, Beijing 100875, China}

\author{Y.~Song}
\affiliation{Center for Correlated Matter and School of Physics, Zhejiang University, Hangzhou 310058, China}

\author{R.~J.~Birgeneau}
\affiliation{Department of Physics, University of California, Berkeley, CA 94720, USA}
\affiliation{Material Sciences Division, Lawrence Berkeley National Lab, Berkeley, California 94720, USA}

%\iffalse
%Since the orbital and Ising-nematic order break the same symmetry, characterizing the nematic and magnetic transitions is crucial for understanding the intertwined nematic, magnetic, and orbital degrees of freedom in iron pnictides and chalcogenides. While the magnetic and nematic phase transitions in hole and electron-doped BaFe$_2$As$_2$ can be attributed to spin fluctuations in a three dimensional lattice with weak out-of-plane anisotropy, a quasi-2D material exhibiting strong anisotropy is still absent, leaving the comprehensive picture incomplete.
%\fi

\date{\today}
\begin{abstract}
Employing inelastic X-ray scattering and neutron scattering techniques, we observed nematic and magnetic phase transitions with distinct characters in K$_{5}$Fe$_{4}$Ag$_{6}$Te$_{10}$. 
%The nematic order undergoes a strongly first-order phase transition, accompanied by a second-order magnetic transition at $T_{\textrm{N}}$~$\approx$~34.6~K. %= 34.62(4)~K. 
%The temperature difference between these two phase transitions is $\sim$~1~K with the nematic one occurring first upon cooling.
Upon cooling, the nematic order undergoes a strongly first-order phase transition followed by a second-order magnetic transition at $T_{\textrm{N}}$~$\approx$~34.6~K. The temperature difference between these two phase transitions is $\sim$~1~K. 
The observed phenomenon can be attributed to a distinctive first-order preemptive Ising-nematic transition, a characteristic unique to a quasi-two-dimensional scenario marked by strong out-of-plane spatial anisotropy due to weak coupling. Our studies establish K$_{5}$Fe$_{4}$Ag$_{6}$Te$_{10}$ as the first material in the family of iron pnictides and chalcogenides that possesses a nematic tricritical point preceding the magnetic one upon decreasing nematic coupling.
\end{abstract}

\maketitle
%Iron pnictides and chalcogenides frequently exhibit ground states that break the $C_{4}$ tetragonal symmetry but with unbroken $O(3)$ spin-rotational symmetry, resembling nematics in liquid crystals~\cite{doi:10.1126/science.1181083, Kasahara2012, doi:10.1126/science.1190482, PhysRevLett.103.057002}. 
\section{Introduction}
Transition metal pnictides and chalcogenides can exhibit intriguing properties including unconventional superconductivity, charge density waves and other correlated states with magnetism~\cite{Paglione2010, RevModPhys.83.1589,Si2016, PhysRevLett.131.186701, Wilson1975, PhysRevLett.122.147601, Bohmer_2018,fernandes_manifestations_2022}. In each of these cases, the interplay between the various order parameters together with their concomitant fluctuations is key to their physics~\cite{Fernandes2014,Fernandes2012,RevModPhys.87.457}. Of particular interest is the sub-class of materials which manifest a ground state that breaks the $C_{4}$ tetragonal symmetry but with unbroken $O(3)$ spin-rotational symmetry, resembling nematics in liquid crystals~\cite{doi:10.1126/science.1181083, doi:10.1126/science.1190482, PhysRevLett.103.057002}. 
The origin of these nematic states is extensively studied as an avenue for comprehending the unconventional superconductivity that commonly arises in the nematic phase~\cite{doi:10.1126/science.aab0103, PhysRevLett.114.157002, PhysRevX.13.011032}. Due to the coupled orbital, magnetic and structural degrees of freedom, experimental identification of the mechanism driving nematic order is challenging~\cite{Fernandes2014}. In this context, a detailed comparison of phase transition behavior in iron pnictides and chalcogenides with various theories serves as a crucial cornerstone for comprehending the underlying nematic driving forces.

%EXPAND THIS PARAGRAPH!!!
%One of the proposed scenarios for the emergence of nematics is fully magnetic, originating from the Ginzburg-Landau effective action for the two low-energy magnetic fluctuations $\Delta_{X}$ and $\Delta_{Y}$. Consequently, an Ising-nematic fluctuation term $g(\Delta_{X}^{2}-\Delta_{Y}^{2})^{2}$ arises through the mean-field approach~\cite{PhysRevB.85.024534}. This model has qualitatively explained the nature of the nematic and magnetic phase transitions in pristine, electron, and hole-doped BaFe$_2$As$_2$ within the framework of a moderate out-of-plane anisotropic perspective, where the magnetic tricritical point precedes the nematic one~\cite{PhysRevB.83.134522, PhysRevB.84.092501}. However, to complete this picture, a material exhibiting strong anisotropy whilst being close to a two-dimensional (2D) spin system is required. This leads to a switching of the nematic and magnetic tricritical points resulting in a first-order nematic transition occurring above the onset of the magnetism.% in the current research of iron pnictides and chalcogenides.

\begin{figure}[t]
    \includegraphics[width=\columnwidth]{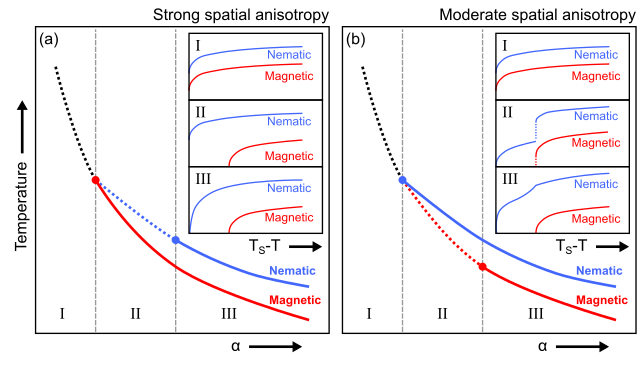}
    \centering
    \caption{Schematic illustration of nematic (blue) and magnetic (red) phase boundaries as a function of the ratio between the magnetic and nematic coupling strength $\alpha \propto g^{-1}$. First-order (second-order) transitions are denoted with a dotted (solid) line and a black dotted line denotes simultaneous first-order nematic and magnetic transitions. (a) shows the case when spatial anisotropy is strong (quasi-2D) and (b) when it is moderate. The tricritical points split each diagram into three regions and the INSETS display typical order parameter behavior for each region. Following~\cite{PhysRevB.85.024534}.}
    \label{Fig_0}
\end{figure}

One of the proposed scenarios for the emergence of nematicity in iron-based systems is fully magnetic, originating from the Ginzburg-Landau effective action for the two low-energy magnetic fluctuations $\Delta_{X}$ and $\Delta_{Y}$. Consequently, an Ising-nematic fluctuation term $g(\Delta_{X}^{2}-\Delta_{Y}^{2})^{2}$ arises through the mean-field approach where $g$ is the nematic coupling strength~\cite{PhysRevB.85.024534}. The main parameters that were found to control the phase transition character (at fixed temperature) were the ratio between the magnetic and nematic coupling strength $\alpha \propto g^{-1}$ and the degree of spatial magnetic anisotropy (i.e. the dimensionality of the system). For systems with fixed but arbitrary dimensionality $2 \leq d \leq 3$, $\alpha$ tunes a crossover between a simultaneous first-order transition in the strong-coupling limit to split second-order magnetic and nematic transitions as $\alpha$ increases. This crossover is manifested by two tricritical points where the nematic and magnetic transitions change from first to second-order respectively as illustrated in Fig.~\ref{Fig_0}.

%For systems with arbitrary dimension $2 < d < 3$, $\alpha$ tunes a crossover between qualitatively three-dimensional (3D) behavior in the strong coupling limit to qualitatively two-dimensional (2D) behavior as $\alpha$ increases. This crossover is manifested by two tricritical points where the nematic and magnetic transitions change from first to second-order respectively.

In a quasi-two-dimensional (quasi-2D) system with weak inter-plane interactions (shown in Fig.~\ref{Fig_0}(a)), the strong-coupling regime (small $\alpha$) exhibits a first-order Ising-nematic transition at $T_{\textrm{S}}$ which is predicted to drive a simultaneous jump in the magnetic order parameter to a finite value. This gives a simultaneous first-order magnetic-nematic transition at $T_{\textrm{N}} = T_{\textrm{S}}$. However, as $\alpha$ increases, the calculated magnitude of the magnetization jump decreases to zero at the first tricritical point. Above this, the first-order Ising-nematic transition at $T_{\textrm{S}}$ persists but the jump in the order parameter is not enough to trigger the magnetism. Hence the magnetic order emerges through a continuous transition at $T_{\textrm{N}} < T_{\textrm{S}}$ as the nematic order parameter grows in the ordered phase. This is shown in Fig.~\ref{Fig_0}(a) - Region II.

%Aspect ratio = 1.701058; PRL word equivalent = 109 words
%Caption = 90 words 

The difference between $T_{\textrm{N}}$ and $T_{\textrm{S}}$ continues to grow as the nematic coupling weakens until the jump of the nematic order parameter also approaches zero. This represents the second tricritical point (the `nematic tricritical point') above which both magnetic and Ising-nematic transitions are second-order (Fig.~\ref{Fig_0}(a) - Region III). Interestingly, the splitting between $T_{\textrm{N}}$ and $T_{\textrm{S}}$, whilst increasing through the nematic tricritical point, eventually decreases to zero as $\alpha \rightarrow \infty$. This nonmonotonic behavior was ascribed to competition between the nematic and magnetic correlation lengths underlying the importance of characterizing both orders.

%The critical values of $\alpha$ can also be influenced by temperature with only a simultaneous first-order transition realised when the transition is in the quantum limit ($T \rightarrow 0$). However, the magnetic tricritical point always precedes the nematic one in a quasi-2D system.

%The position of the tricritial points is also dependent on the temperature with only a simultaneous first-order transition realised in the quantum limit ($T \rightarrow 0$). However, for all control parameters ($\alpha$, temperature, dimensionality) the magnetic tricritical point always precedes the nematic one in a quasi-2D system.

%This is not the case when the out-of-plane anisotropy is moderate. $\alpha$ still tunes a crossover between simultaneous first-order magnetic and Ising-nematic transitons to split second-order transitions but the order of the nematic and magnetic tricritical points was found to be reversed. By parameterizing the anisotropy by $\eta$ which represents coupling between layers of the crystal, it was found that tricritical points cross when $\eta \sim 0.43$ ($\eta = 0$ corresponds to the fully 2D case). The moderate-anisotropic limit was also found to be able to support metanematic transitions where, despite displaying a second-order Ising-nematic transition at $T_{\textrm{S}}$, the nematic order parameter jumps coincident with $T_{\textrm{N}}$. 

This is not the case when the spatial out-of-plane anisotropy is moderate shown in Fig.~\ref{Fig_0}(b). $\alpha$ still tunes a crossover between simultaneous first-order magnetic and Ising-nematic transitons to split second-order transitions but the order of the nematic and magnetic tricritical points was found to be reversed. Specifically, considering a stacked 2D layered material and parameterizing the anisotropy by $\eta < 1$ representing inter-layer coupling, the inverse magnetic susceptibility can be written as $\chi_{i,q}^{-1} \sim r_0 + q^2_{\parallel} + \eta q_z^2$ where $i$~=~X or Y, $r_0$ is the quadratic coefficient in the effective action and $\parallel$ and $z$ correspond to in-plane and out-of-plane respectively. It was found that tricritical points swap position with respect to $\alpha$ when $\eta \sim 0.43$ ($\eta = 0$ corresponds to the fully 2D case). The moderate-anisotropic limit was also found to be able to support metanematic transitions where, despite displaying a second-order Ising-nematic transition at $T_{\textrm{S}}$, the nematic order parameter jumps coincident with $T_{\textrm{N}}$ (Fig.~\ref{Fig_0}(b) - Region II). 

This framework has qualitatively explained the nature of the nematic and magnetic phase transitions in 122 systems such as CaFe$_2$As$_2$ and SrFe$_2$As$_2$ where a simultaneous first-order nematic-magnetic transtions was observed~\cite{PhysRevB.79.224518,PhysRevB.78.014523}. It also can be applied to pristine, electron, and hole-doped BaFe$_2$As$_2$~\cite{PhysRevB.83.134522, PhysRevB.84.092501,PhysRevB.80.060501,annurev070909-104041}. Indeed, BaFe$_2$As$_2$ has also been shown to display a broadening of the Bragg peak prior to a first-order structural transition which is consistent with the metanematic transition predicted in the moderate-anisotropy, intermediate-coupling case (Region II in Fig.~\ref{Fig_0}(b))~\cite{PhysRevB.82.144525}. However, the small splitting implies that $\alpha$ is small in pristine BaFe$_2$As$_2$. Co doping of BaFe$_2$As$_2$ adds carriers to the electron pockets and so increases the chemical potential. This is predicted to increase $\alpha$ making electron doped compounds more likely to exhibit split magnetic and nematic transitions. Ba(Fe$_{1-x}$Co$_x$)$_2$As$_2$ shows a splitting of the magnetic and nematic transitions as $x$ increases with both becoming second-order for $x > 0.022$ in agreement with this theory~\cite{PhysRevB.83.134522,PhysRevB.84.092501}.

However, to complete this picture, a material exhibiting strong real-space anisotropy whilst being close to a 2D spin system is required. In this work, we present a comprehensive investigation into the phase transitions of the iron chalcogenide, K$_{5}$Fe$_{4}$Ag$_{6}$Te$_{10}$ (KFAT). Employing inelastic X-ray and neutron scattering techniques, we have elucidated the characteristics of both the magnetic and nematic-coupled structural transition order parameters below the transition temperature, along with their corresponding fluctuations. Our findings reveal a strongly first-order nematic transition and a second-order magnetic transition in KFAT, a phenomenon unprecedented in other iron chalcogenides or pnictides. This is an unexpected result and can likely be attributed to a scenario of local magnetism, where the magnetic order is induced by the spin-nematic phase.

The growth and physical properties of KFAT single crystals have been described previously~\cite{PhysRevB.84.060506}. Neutron scattering measurements were carried out on the triple-axis spectrometer HB-3 (TAX), HIR, Oak Ridge National Laboratory. Single-crystal elastic and inelastic X-ray scattering data were collected at BL35XU, SPring-8~\cite{BARON2000461}. Experimental details are described in the Supplemental Material~\cite{SM}. %For simplicity and clarity, all Bragg peaks in this letter will be indexed in the tetragonal system noted with a subscript `T', following the BaFe$_2$As$_2$ convention.

\nocite{baron2020,PhysRevB.12.368,OLIVERO1977233}

%Aspect ratio = 0.94731; PRL word equivalent = 179 words
%Caption = 155 words 
\begin{figure}[t]
    \includegraphics[width=\columnwidth]{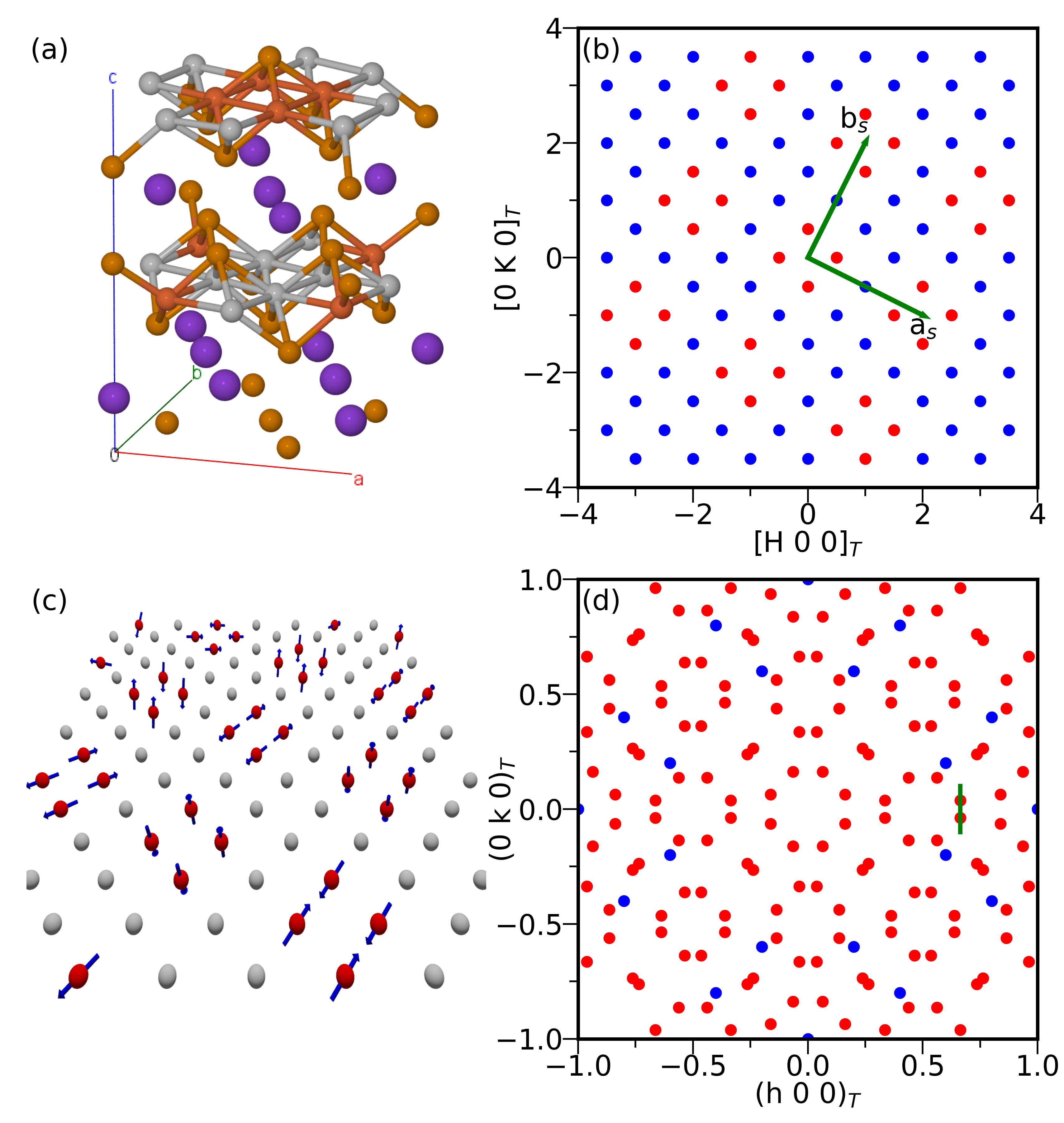}
    \centering
    \caption{Crystal and magnetic structure of KFAT. (a) Crystal structure of KFAT: purple, yellow, grey and orange spheres represent the potassium, tellurium, silver and iron atoms, respectively. (b) The Fe-Ag plane of KFAT. The red dots represent the iron atoms and blue dots represents the silver atoms. The iron clusters are arranged in the $\sqrt{5}\times\sqrt{5}$ pattern. The superstructure lattice vectors are labeled. (c) The magnetic structure of KFAT is characterized by a collinear antiferromagnetic arrangement within each iron cluster, resembling the structure observed in BaFe$_2$As$_2$. However, between these clusters, there is a uniform rotation of the spins, resulting in an incommensurate order. (d) Structural (blue) and magnetic peak (red) locations within the (hk0)$_{\textrm{T}}$ reciprocal space. The plotted locations contains peaks from all possible structural magnetic domains from a large single crystal used for neutron scattering experiment. Critical scattering scans were performed along the green lines shown in the figure.}
    \label{Fig_1}
\end{figure}

\section{Results and Discussion}
\subsection{Critical Behavior of Magnetism}

Previous studies indicate that KFAT shares similarities with tetragonal BaFe$_2$As$_2$~\cite{PhysRevLett.122.087201}, as shown in Fig.~\ref{Fig_1}(a). However, unlike in electron-doped BaFe$_2$As$_2$ where the dopant distribution can usually be taken as random, 
%Unlike the random distribution of iron atoms on the square lattice in electron-doped BaFe$_2$As$_2$,
in KFAT, the iron atoms organize into $2\times2$ clusters arranged in a $\sqrt{5}\times\sqrt{5}$ expansion of the tetragonal unit cell, with silver atoms serving as inter-cluster separators, as Fig.~\ref{Fig_1}(b) presents. For simplicity and clarity, all Bragg peaks in this paper will be indexed in the tetragonal system noted with a subscript `T', following the BaFe$_2$As$_2$ convention. The spatial separation of iron atoms restricts electron hopping between clusters, classifying KFAT as a semiconductor with a narrow band gap~\cite{PhysRevB.84.060506}, indicative of localized magnetism. Meanwhile, the superexchange facilitated by silver atoms remains influential in inter-cluster magnetic coupling. Their competition with direct exchange within the iron cluster results in a unique incommensurate magnetic structure characterized by an ordering vector within the Fe-Ag plane below $T_{\textrm{N}}$~=~35~K (Fig~\ref{Fig_1}(c)). The corresponding structural and magnetic peak locations are labeled as blue and red dots in Fig~\ref{Fig_1}(d), respectively. A nematic transition occurs nearly concurrently, as evidenced by the coupled structural distortion, manifested in the broadening of Bragg peaks observed during neutron diffraction experiments~\cite{PhysRevLett.122.087201} (also see Supplementary Material~\cite{SM}). The refined magnetic moments on the iron atoms in KFAT are found to be 2.11(3)~$\mu_{B}$~\cite{PhysRevLett.122.087201}, making it an ideal material for exploring magnetic fluctuations above the ordering temperature. This is particularly significant when compared with BaFe$_2$As$2$, where the magnetic moments are approximated to be 0.87(3)~$\mu_{B}$~\cite{PhysRevLett.101.257003}. 

%Aspect ratio = 1.00559; PRL word equivalent = 170 words
%Caption = 122 words 
\begin{figure}[t]
    \includegraphics[width=\columnwidth]{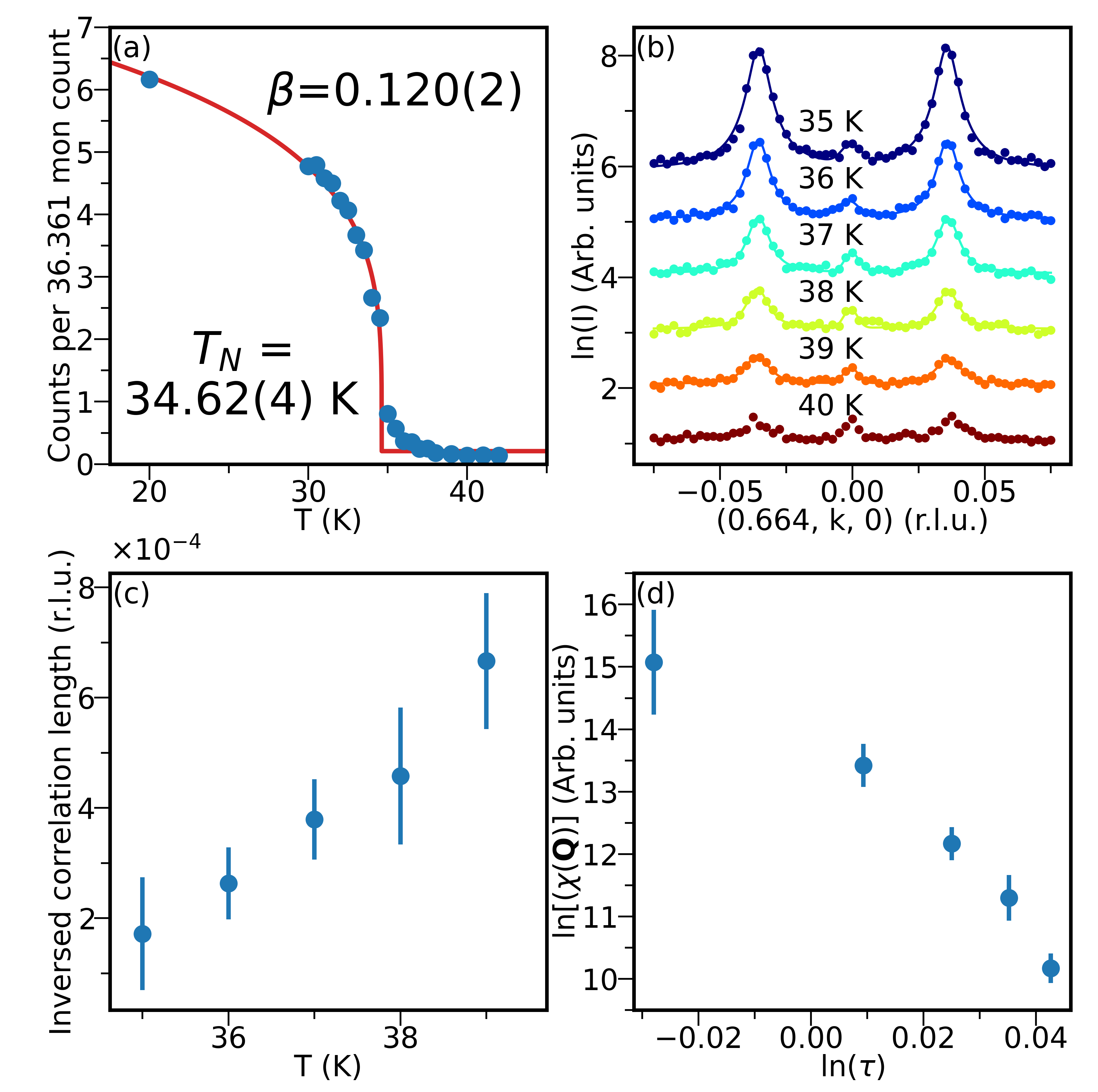}
    \centering
    \caption{Behavior of magnetism in KFAT. (a) temperature dependence of intensity at $\textbf{Q}$~=~(0.664,~0.038,~0)$_{\textrm{T}}$. The red curve shows the order parameter fit. (b) Critical scattering scans along (0.664,~k,~0) above the N{\'e}el temperature $T_{\textrm{N}}$ in logarithmic scale. From top to bottom are data acquired at T = 35, 36, 37, 38, 39, and 40~K, respectively. The corresponding fits are plotted with solid lines except for 40 K due to statistical insufficient counts. (c) Inverse correlation length in reciprocal lattice units (r.l.u.), obtained via critical scattering fitting, convoluted with the resolution function. (d) Logarithmic plot of critical scattering intensity~$\propto$~$\chi(\textbf{Q})$ against reduced temperature $\tau=(T-T_{\textrm{N}})/T_{\textrm{N}}$. The correlation length and critical scattering strength at 40~K is not presented because the model does not give a converged result.}
    \label{Fig_2}
\end{figure}

Figure~\ref{Fig_2}(a) shows the temperature dependence of the magnetic Bragg peak intensity at $\textbf{Q}$~=~(0.664,~0.038,~0)$_{\textrm{T}}$. An effective critical exponent of $\beta = 0.120(2)$ was extracted from the order parameter fit using data ranging from $T = 30$ K to $T = 42$ K, with an onset of magnetic ordering temperature $T_{\textrm{N}} = 34.62(4)$~K. The obtained order parameter critical exponent is consistent with that measured in a 2D Ising systems such as K$_2$CoF$_4$ and Rb$_2$CoF$_4$~\cite{SAMUELSEN1974785, IKEDA1974529}. However, in the case of these fluorides, their 2D nature can be readily understood due to an inter-plane cancellation resulting in much larger relative in-plane coupling~\cite{PhysRevB.1.2211, PhysRevB.3.1736}. Furthermore, the magnetic moments are pointed along the c axis making the 2D Ising interpretation a natural description. This is unexpected in KFAT as the magnetic structure consists of rotating spins and so should not generally be considered to be Ising-like. Though the exact Hamiltonian determination should be based on information about spin dynamics obtained through inelastic neutron scattering, the scope of the discussion here will be limited to the 2D Ising model in order to be consistent with the effective critical exponent of the order parameter.

To delve deeper into the magnetic transition in KFAT, transverse scans were conducted along (0.664,~k,~0)$_{\textrm{T}}$ along the green lines in Fig.~\ref{Fig_1}(d) and above $T_{\textrm{N}}$ to acquire critical scattering data. These scans are presented in Fig.~\ref{Fig_2}(b). The transverse instrument resolution is determined by the rocking scans across nuclear peaks at \textbf{Q} = (0.6,~0.2,~0)$_{\textrm{T}}$, (0.6,~-0.2,~0)$_{\textrm{T}}$, and (0.2,~0.6,~0)$_{\textrm{T}}$. The nuclear peak profiles of these scans can be described by a Lorzentian function. They provide a good approximation of the transfer resolution function at (0.664,~k,~0)$_{\textrm{T}}$ due to similar scattering angles and the absence of observable Bragg peak splitting below the nematic transition temperature. The longitudinal resolution function is Gaussian and obtained via $\theta$-2$\theta$ scans across the magnetic peak at 4~K. As a result, the line-shape of critical scattering in Fig.~\ref{Fig_2}(b) can be fit by a convolution of the resolution function and 2D critical scattering profile described by a 2D Lorentzian function (see Supplementary Material~\cite{SM}). %Note that the measured critical scattering, by necessity, involves a combination of transverse and longitudinal (to the spin direction) fluctuations but only the latter diverge. The obtained inverse correlation lengths are presented in Fig.~\ref{Fig_2}(c). Extrapolation of the fitted correlation length shows that it reaches a finite value at the transition temperature. This is a unique feature of the 2D Ising model, where the correlation length of transverse spin fluctuations does not diverge~\cite{PhysRevB.16.280}. Further evidence is provided by the strength of the critical fluctuations \(\chi_{\textbf{Q}}\) illustrated in Fig.~\ref{Fig_2}(d). These fluctuations deviate from the phenomenological scaling relation due to the presence of transverse spin fluctuations which do not follow the scaling relation. A refined model involving both longitudinal and transverse spin fluctuations does not yield converged fitting results, because of the resolution-limited magnetic Bragg diffraction in the critical region. Nevertheless, the critical behavior of correlation length and fluctuation strength supports the presence of critical scattering and is consistent with the second-order nature of the magnetic transition.
Note that the measured critical scattering, by necessity, involves a combination of fluctuations parallel and perpendicular to the spin direction but only the former diverge. The obtained inverse correlation lengths are presented in Fig.~\ref{Fig_2}(c). Extrapolation of the fitted correlation length shows that it reaches a finite value at the transition temperature. This is a unique feature of the 2D Ising model, where the correlation length of perpendicular spin fluctuations does not diverge~\cite{PhysRevB.16.280}. Further evidence is provided by the strength of the critical fluctuations \(\chi_{\textbf{Q}}\) illustrated in Fig.~\ref{Fig_2}(d). These fluctuations deviate from the phenomenological scaling relation due to the presence of perpendicular spin fluctuations which do not follow the scaling relation. A refined model involving both parallel and perpendicular spin fluctuations does not yield converged fitting results, because of the resolution-limited magnetic Bragg diffraction in the critical region. Nevertheless, the critical behavior of correlation length and fluctuation strength supports the presence of critical scattering and is consistent with the second-order nature of the magnetic transition.

%Aspect ratio = 1; PRL word equivalent = 170 words
%Caption = 176 words
\begin{figure}[t]
    \includegraphics[width=.9\linewidth]{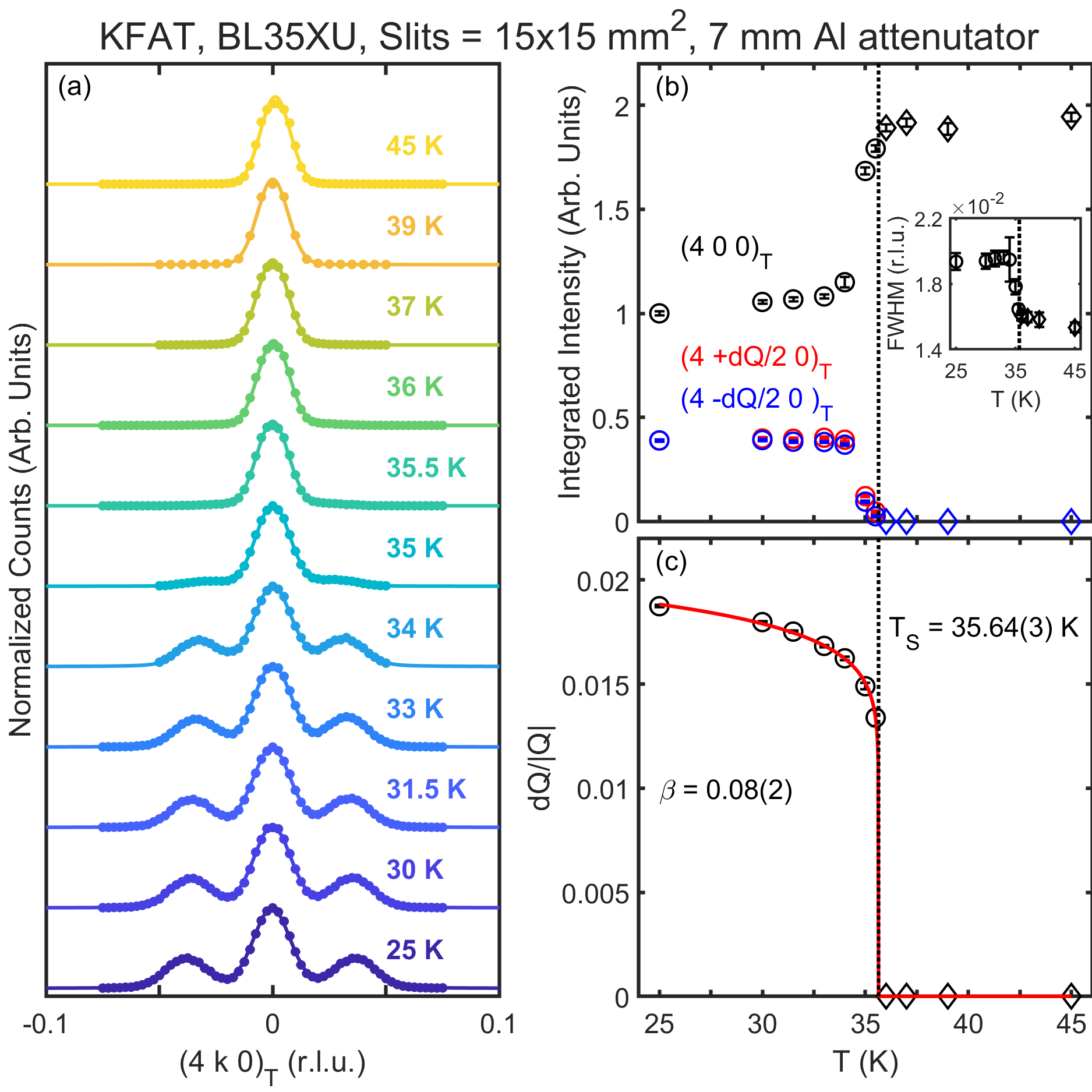}
    \centering
    \caption{Nematic transition in KFAT. (a) Elastic \textbf{Q} scans across the (4,~0,~0)$_{\textrm{T}}$ Bragg peak. All scans are normalized such that the central peak amplitude is the same to emphasize the change in the peak profile. Below the nematic transition, the data were fitted using three Voigt functions constrained to have the same FWHM but independent amplitudes. The intensity is plotted in (b). The fitted FWHM (INSET) is comparable to the instrument \textbf{Q} resolution ($\sim 0.019$~r.l.u. along k) and so remains resolution-limited at all temperatures. For $T \geq 36$~K we observed no splitting within our resolution so these points are set to zero and are denoted by diamond markers (the \textbf{Q} scan was only fitted with one Voigt component). The peak splitting $|$d$\mathbf{Q}|$ is plotted in (c) normalized by $|\mathbf{Q}|$ which may be treated as a de-facto order parameter (see discussion in main text). The quick saturation of the satellite intensity suggests a strongly first order transition. The vertical dotted line in (b) and (c) denotes the fitted $T_{\textrm{S}} = 35.64(3)$~K. }
    \label{Fig_3}
\end{figure}

\subsection{Critical Behavior of Lattice}

To investigate the nematic transition, we employed X-ray scattering techniques. Both elastic and inelastic data presented here were taken on the same sample on beamline BL35XU at SPring-8 during a single experiment. Unlike typical X-ray diffraction experiments, which correspond to energy-integrated scattering and include thermal fluctuations over a broad temperature range, BL35XU offers energy-resolved measurements with a measured full width at half maximum (FWHM) of $\sim$1.58~meV for our experiment. This allows us to effectively integrate only over thermal fluctuations within a narrow range of approximately 20~K. Fig.~\ref{Fig_3}(a) shows transverse elastic \textbf{Q} scans across the (4,~0,~0)$_{\textrm{T}}$ Bragg peak upon cooling. Due to the coupling to the lattice, the nematic order splits this Bragg peak with satellites at (4,~$\pm|$d$\mathbf{Q}|$/2,~0)$_{\textrm{T}}$. This arises from the loss of the four-fold symmetry which generates four structural domains below $T_{\textrm{S}}$. Each satellite peak of (4, 0, 0)$_{\textrm{T}}$ corresponds to a single domain with shear distortion of lattice vector $\mathbf{a}_{\textrm{T}}$, while the remaining two structural domains with shear distortion of $\mathbf{b}_{\textrm{T}}$ both contribute to the central Bragg peak. The \textbf{Q} scans at each temperature were fitted by three Voigt profiles with independent amplitudes but the same FWHM. For $T \geq 36$~K no satellites are seen and the data were fitted with a single Voigt component. As Figs.~\ref{Fig_3}(a) and~\ref{Fig_3}(b) show, the intensity of satellite peaks rise sharply at $\approx$~35~K without any prior notable broadening of the central Bragg peak (fitted FWHM shown in the inset of Fig.~\ref{Fig_3}(b) suggests that the peak is resolution-limited). This is in contrast to BaFe$_2$As$_2$ where a preemptive second-order structural transition characterized by a broadening of the Bragg peak appears before the first-order (metanematic) transition~\cite{PhysRevB.82.144525}. Whilst a finer temperature steps would be required to discount this absolutely conclusively in KFAT, the lack of phonon softening (discussed below) would suggest that this transition is more first-order, again in contrast with BaFe$_2$As$_2$~\cite{niedziela_2011}.

The temperature evolution of the splitting, serving as an indicator of the nematic order parameter, is illustrated in Fig.~\ref{Fig_3}(c). The shear structural distortion causes a difference in $a_{O}$ and $b_{O}$, which is the spacing between Fe/Ag atoms along [110]$_{\textrm{T}}$ and [1$\bar{1}$0]$_{\textrm{T}}$ directions respectively and is presented in Fig.~\ref{Fig_1}(b). Thus, this shear distortion results in a monoclinic structure, which is commonly referred to as orthorhombic distortion in the literature of Fe-based pnictides/chalogenides~\cite{PhysRevLett.101.257003, PhysRevB.79.180508}. For the transverse scan across (4,~0,~0)$_{\textrm{T}}$, $|\rm{d}\mathbf{Q}|/|\mathbf{Q}| \approx 4\delta$ where $\delta = (a_{O}-b_{O})/(a_{O}+b_{O})$. To a first order approximation, $\delta$ is proportional to the shear angle. Our fitting gives $\delta \approx 5.2(3)\times 10^{-3}$ which roughly agrees with the literature value for KFAT ($\approx 3.8 \times 10^{-3}$ measured with neutrons~\cite{PhysRevLett.122.087201}) and is close to that measured in BaFe$_{2}$As$_2$ ($\approx 4 \times 10^{-3}$)~\cite{PhysRevLett.101.257003}. The reported value of \(\delta\) in this work is more accurate than that in Ref.~\onlinecite{PhysRevLett.122.087201}, thanks to the fully resolved Bragg peak splitting with synchrontron X-ray measurements. To examine further the nature of the structural transition, we attempted an order parameter fitting for the splitting. This gives a structural transition temperature $T_{\textrm{S}} = 35.64(3)$~K which is $\sim$~1~K above $T_{\textrm{N}}$. %The extracted effective critical exponent $\beta = 0.08(2)$ is potentially influenced by sample inhomogeneity of transition temperatures~\cite{10.1063/1.1456441} or temperature evolution of order parameter below $T_{\textrm{S}}$.
The extracted effective critical exponent $\beta = 0.08(2)$ is potentially influenced by sample inhomogeneity resulting in a distribution of transition temperatures~\cite{10.1063/1.1456441} or temperature evolution of order parameter below $T_{\textrm{S}}$ but the small value of $\beta$ is suggestive of an underlying first-order transition.

One hallmark of a second-order phase transition is the emergence of softening in the corresponding fluctuations prior to their condensation. Indeed, in electron and holed-doped BaFe$_2$As$_2$, previous studies have shown softening in the in-plane transverse acoustic (IPTA) mode above the critical temperature~\cite{PhysRevB.98.014516}. In particular, the dispersion of the IPTA phonon in the vicinity of the gamma point can be described by mean-field theory with the weak coupling between lattice and nematicity by the following formula

\begin{equation}%PRL word equivalent = 16 words
    \frac{E_{\xi}(k)}{E_{\xi=0}(k)}=\sqrt{\frac{1+\xi^{2}k^{2}}{1+\xi^{2}(k^{2}+r)}},
\end{equation}

\noindent where $\xi$ is the nematic correlation length and $r$ is a temperature independent parameter~\cite{PhysRevLett.124.157001}. Supposing KFAT undergoes a similar second-order nematic transition, one would expect a divergence of the nematic correlation length, leading to softening of the IPTA phonon and its related shear modulus $c_{66}$~\cite{PhysRevLett.105.157003, doi:10.1143/JPSJ.81.024604}. 

%Aspect ratio = 3/4; PRL word equivalent = 220 words
%Caption = 144 words
\begin{figure}[t!]
    \includegraphics[width=.9\linewidth]{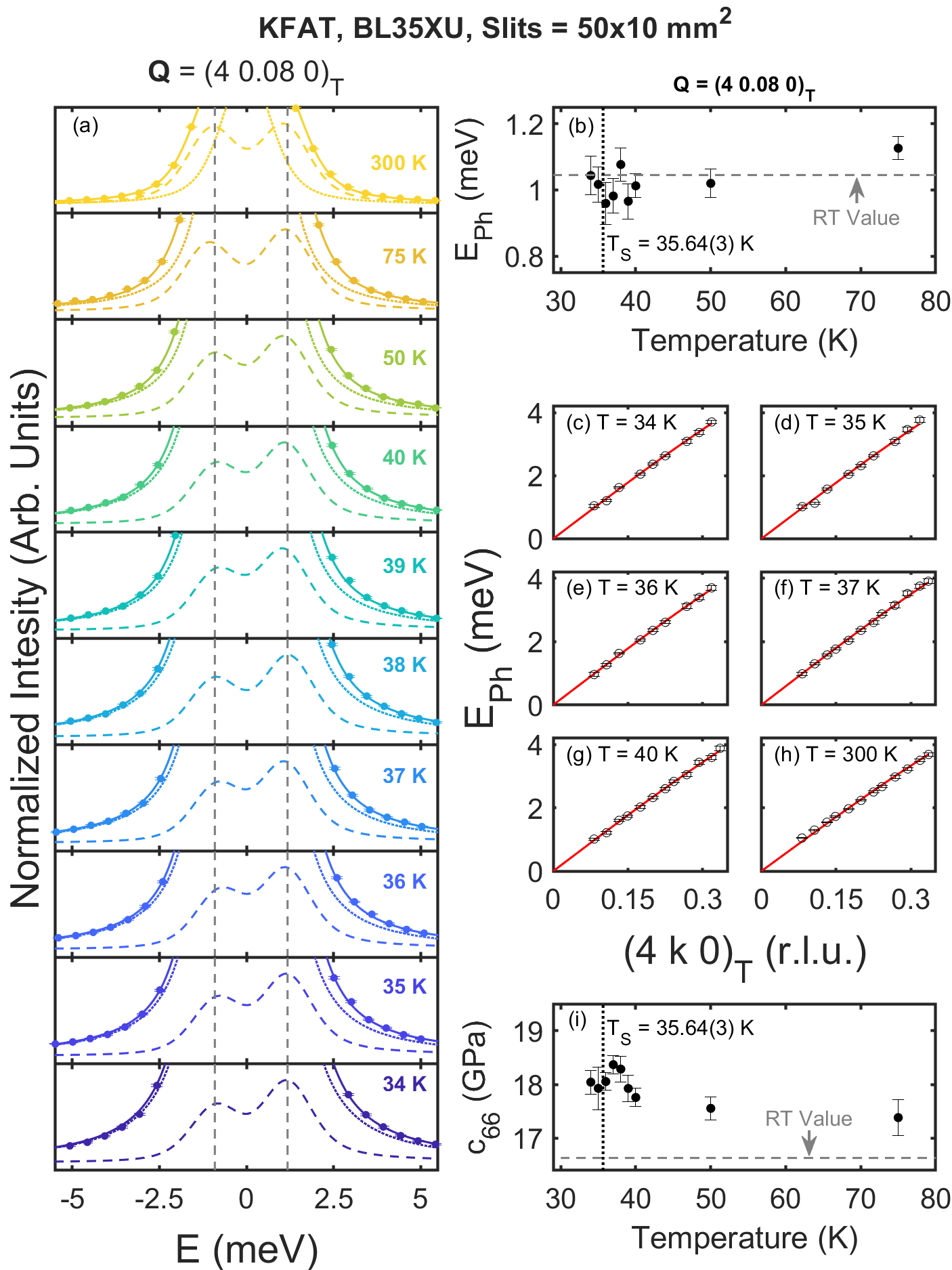}
    \centering
    \caption{Inelastic X-ray scattering in KFAT. Representative energy scans at $\mathbf{Q}$~=~(4,~0.08,~0)$_{\textrm{T}}$ are shown in (a) for all temperatures. The colored dotted, dashed, and solid lines denote the quasi-elastic, phonon and total fits respectively. Due to the large quasi-elastic tail, the axes are scaled to emphasize the fitted phonon component at each temperature. The corresponding temperature dependence of the phonon energy at (4,~0.08,~0)$_{\textrm{T}}$ is shown in (b). Fits of the in-plane transverse acoustic dispersion are shown in (c) - (h) for temperatures near $T_{\textrm{S}}$ and room temperature (RT). (i) shows the extracted elastic constant $c_{66}$ from fitted sound velocity. In all panels, gray dashed lines also denote the RT values to aid comparison. Softening of the phonon near $T_{\textrm{S}}=35.64(3)$~K is absent within experimental error with the dispersion being essentially temperature independent.}
    \label{Fig_4}
\end{figure}

Figure~\ref{Fig_4} presents the results of our inelastic X-ray scattering experiment. We measured the dynamical structural factor dominated by the IPTA phonon along (4,~k,~0)$_{\textrm{T}}$ (0.08~$<$~k~$<$~0.34) from room temperature down to 34~K. The phonon energies at fixed \textbf{Q} were determined through fitting energy scans. The fitting model describes the inelastic phonon line shape using Voigt functions, convolved with the measured experimental resolution function and weighted by the Bose factor~\cite{PhysRevLett.126.107001}. Figure~\ref{Fig_4}(a) displays the energy scans at \textbf{Q}=(4,~0.08,~0)$_{\textrm{T}}$, which is the closest Q-vector to $\Gamma$ with a resolvable phonon peak and, therefore, should display the strongest nematic-induced phonon softening. However, the change in the IPTA phonon energy (Fig.~\ref{Fig_4}(b)) upon cooling is not statistically significant being within fitted error of phonon energy. Our analysis shows no IPTA phonon softening during the nematic phase transition, ruling out the possibility of a second-order nematic/structural transition. Yet it is consistent with expectations for a first-order phase transition where softening is absent due to the nematic correlation length not significantly changing. We further extract the IPTA phonon dispersion to examine our speculation of the strongly first-order phase transition. This is shown in Figs.~\ref{Fig_4}(c)-(h) and the data were fitted using $E_{\textrm{Ph}}(k) = A \sin(D \pi k)/(D \pi)$ with $D=0.5$ which adequately reproduces the data. We find the IPTA phonon does not exhibit any notable temperature dependence from room temperature down to 34~K and could be described by a Debye model. The obtained shear modulus $c_{66}$ is plotted in Fig.~\ref{Fig_4}(i) and shows a gradual temperature dependence but remains constant across T$_{S}$ within the fitted error, serving as another indicator of the strongly first-order nematic phase transition in KFAT.

%\section{discussion}
\section{Conclusions}

In conclusion, we report evidence of a first-order structural and second-order quasi-2D Ising magnetic transition in KFAT. Despite the quasi-2D crystal structure, this type of coexistence has not been detected in iron pnictides and chalcogenides before. Apart from experimental challenges due to the small magnetic moments, previous studies indicate the transition of BaFe$_2$As$_2$ magnetic fluctuations from 2D to 3D~\cite{PhysRevB.79.184519, PhysRevB.82.144502}, leaving BaFe$_2$As$_2$ in a moderately spatially anisotropic state where the nematic tricritical point was approached after the magnetic one with increasing $\alpha$. This could result from the interlayer exchange coupling of the Fe spins. In KFAT, the in-plane projections of iron clusters in the neighboring planes are staggered, which increases the nearest Fe-Fe bond length between layers and significantly weakens the interlayer exchange coupling. This unique staggered Fe cluster configuration results in quasi-2D magnetism and drives KFAT into a strong spatially anisotropic scenario, where a first-order preemptive Ising-nematic transition is allowed~\cite{PhysRevB.85.024534}.

One of the unique properties of the quasi-2D Ising-nematic transition predicted by mean-field theory is the occurrence of a pseudo-gap phase~\cite{PhysRevB.85.024534}. This pseudo-gap phase partially gaps the Fermi surface and is generic in cuprates~\cite{RevModPhys.78.17, Timusk_1999}. It arises due to the magnetic correlation length jumping with the enhancement of thermal-magnetic fluctuations in the quasi-2D scenario. As a result, the electronic spectra develop a magnetic pseudogap through the transfer of its spectral weight, although zero-frequency states appear only below $T_{N}$. Though the observed magnetic critical scattering is highly resolution-limited in the critical region, rendering such analysis not practical, we would like to note the potential relation between the semiconductor nature of KFAT and pseudogap phase.

\begin{acknowledgements}
The work at UC Berkeley and LBNL was funded by the U.S. Department of Energy, Office of Science, Office of Basic Energy Sciences, Material Sciences and Engineering Division under Contract No. DE-AC02-05-CH11231 (Quantum Materials program KC2202). A portion of this research used resources at the High Flux Isotope Reactor, a DOE Office of Science User Facility operated by the Oak Ridge National Laboratory. The synchrotron radiation experiments were performed at BL35XU of SPring-8 with the approval of the Japan Synchrotron Radiation Research Institute (JASRI) under Proposal No 2023B1501. The work at Beijing Normal University is supported by the Fundamental Research Funds for the Central Universities (Grant No. 2243300003).
\end{acknowledgements}

\bibliography{apssamp}% Produces the bibliography via BibTeX.

\end{document}